# Glass transitions, semiconductor-metal (SC-M) transitions and fragilities in Ge-V-Te (V=As, or Sb) liquid alloys: the difference one element can make

Shuai Wei[1,2], Garrett J. Coleman[2], Pierre Lucas[2], and C. Austen Angell[1*]

[1]*School of Molecular Sciences, Arizona State University, Tempe, AZ 85287-1604, USA*

[2]*Department of Materials Science and Engineering, University of Arizona, Tucson, Arizona 85712, USA*

## ABSTRACT

Glass transition temperatures ($T_g$) and liquid fragilities are measured along a line of constant Ge content in the system Ge-As-Te, and contrasted with the lack of glass-forming ability in the twin system Ge-Sb-Te at the same Ge content. The one composition established as free of crystal contamination in the latter system shows a behavior opposite to that of more covalent system. Comparison of $T_g$ vs bond density in the three systems Ge-As-chalcogen differing in chalcogen i.e. S, Se, or Te, shows that as the chalcogen becomes more metallic, i.e. in the order S<Se<Te the bond density effect on $T_g$ becomes systematically weaker, with a crossover at <r> = 2.3. When the more metallic Sb replaces As at <r> greater than 2.3, incipient metallicity rather than directional bond covalency apparently gains control of the physics. This leads us to an examination of the electronic conductivity and, then, semiconductor-to-metal (SC-M) transitions, with their associated thermodynamic manifestations, in relevant liquid alloys. The thermodynamic components, as seen previously, control liquid fragility and cause fragile-to-strong transitions during cooling. We tentatively conclude that liquid state behavior in phase change materials (PCMs) is controlled by liquid-liquid (SC-M) transitions that have become submerged below the liquidus surface. In the case of the Ge-Te binary, a crude extrapolation to GeTe stoichiometry indicates that the SC-M transition lies about 20% below the melting point, suggesting a parallel with the intensely researched "hidden liquid-liquid (LL) transition", in supercooled water. In the water case, superfast crystallization initiates in the high fragility domain some 4% *above* the $T_{LL}$ which is located at ~15% below the (ambient pressure) melting point.

---

*Author to whom correspondence should be addressed. email: austenangell@gmail.com

## I. INTRODUCTION

Chalcogenide glasses, comprised of chalcogen elements (S, Se, Te) covalently bonded with Group IV and V metalloid elements with similar electronegativity such as As, Sb, Ge and Si, offer attractive optical and electronic properties for potential applications such as infrared optical fibers[1] and phase-change non-volatile memory devices[2,3]. The knowledge of glass transition behavior and kinetic properties of the supercooled liquid state are crucial for applications and the understanding of this class of materials, as they are related to structural relaxations, aging, stability, processability, crystallization kinetics[4], etc. In particular, the temperature dependence of viscosity plays an important role in nucleation and growth rates of crystals and is receiving increasing attention in the field of phase-change materials[5].

The concept of liquid fragility has been found useful in discussion of the wide variety of viscosity behavior observed in glassforming liquids[6]. A near-Arrhenius rise in viscosity, with the "universal" pre-exponential factor $10^{-4}$-$10^{-5}$ Pa·s, is classified as "strong" behavior, whereas a range of non-Arrhenius to highly non-Arrhenius behavior is classified as "fragile". Fragility can be represented by the slope of the $T_g$-scaled Arrhenius-plot (i.e. fragility-plot) for viscosity η (or structural relaxation time τ) at $T=T_g$, so-called "steepness index" or "$m$-fragility",

$$m = \frac{d\log\eta}{d(T_g/T)}\bigg|_{T=T_g}, \qquad (1)$$

where $T_g$ is defined as the temperature where the viscosity increases to the value $\eta = 10^{12}$ Pa·s associated with the rigid behavior[6]. Alternatively, and more commonly, $T_g$ can be determined by differential scanning calorimetry (DSC), as the onset temperature for the jump in heat capacity which occurs as the liquid degrees of freedom are accessed during a "standard DSC scan" (i.e. heating at 20Kmin$^{-1}$ after vitrification at 20 K/min$^{-1}$ ($q_h=q_c=20$ K min$^{-1}$). This glass transition corresponds to the temperature at which the liquid enthalpy relaxation time reaches 100 s)[6]. The two definitions are *not* always the same, as in the case of fragile liquids, the viscosity at $T_g$ tends to be less than $10^{12}$ Pa·s (see Figure 1 of ref.[7]). The $m$ values range from 16 (perfect Arrhenius[8–10] or strong) to $m \approx 170$ for the most fragile liquids[8,11] and 200 if polymers are included.

Determination of fragility usually requires a stable supercooled liquid region, in which the viscosity or structural relaxation times can be measured as a function of temperature[12,13]. However, for poor glass-formers, such as phase-change materials $Ge_2Sb_2Te_5$ (GST225), $Ge_1Sb_2Te_4$ (GST124) and $Ag_4In_3Sb_{67}Te_{26}$ (AIST), the glass transition and supercooled liquid regime are difficult to probe directly, as crystallization occurs on very short timescales and preempts the $T_g$ [14,15]. Thus, indirect methods that employ related quantities such as crystal growth rates have been used to derive the viscosity. For example, Orava et al.[4] studied the crystallization kinetics using nano-calorimetry with extremely high heating rates and estimated the fragility of GST225 to be high, $m \approx 90$. Salinga et al.[5] measured the crystal growth velocity of AIST using time-resolved laser reflectivity experiments. and derived an extremely high fragility of $m = 128$ while Zalden et al.[16], observing crystal growth after femtosecond pulse optical excitation, reported m = 104 close to $T_m$ for the same composition. Greer and coauthors[17] suggest a broad crossover from m = 37 to m = 74 in the same liquid over a wider temperature range. High fragilities near the melting point can generally be expected from the Johnson[18] and Greer[19] methods based on the Turnbull parameter and the crystallization time at the nose of the TTT curve.

On the other hand, there are "good" chalcogenide glass-formers such as Ge-As-Se and Ge-As-S, that are characterized by low-to-intermediate fragilities[20–22] ($m$=27-65). A minimum fragility has been observed at an average bond coordination number (or bond density), $<r>$, higher than the rigidity percolation threshold at $<r> = 2.4$[21,22]. Recently, the fragility of $Ge_{15}Te_{85}$ near $T_g$ has been determined to be relatively strong compared to its high-temperature fragile liquid state[23]. A fragile-to-strong liquid transition was revealed by the Adam-Gibbs equation[24] fitting of viscosity data[23]. It appears that the fragility of chalcogenide liquids can vary over a broad range from as strong as silica[8] ($m$=20) to as fragile as the simple ionic glass-former calcium potassium nitrate CKN ($m$=93, ref.[8]). However, the origin of such diverse liquid dynamics in this Group IV-V-VI class of materials remains unclear, and systematic studies are desirable.

The Ge-As-Te alloys have composition ranges that are good glass-formers[25] with supercooled liquid regions accessible to differential scanning calorimetry (DSC) studies. They differ from Ge-As-Se and Ge-As-S only by the difference in chalcogen elements (Te, Se, S), whereas they differ from Ge-Sb-Te by having a different Group V metalloid. In this regard, Ge-As-Te is a link between former and latter types of alloys. Understanding the analogies and difference between Ge-As-Te and the related alloys may provide new insights into the origin of the diverse fragilities and glass transition behaviors in chalcogenide glasses and help reach a better understanding of the related Ge-Sb-Te phase-change materials[26]. The urgency of understanding the underlying physics of phase change materials in the context of technological applications should need no emphasis.

In this work, we investigate the glass transition, Turnbull parameter, and fragility using DSC on small samples that fall along the line of increasing As (or Sb) atoms replacing Te atoms (Fig. 1). The $T_g$ and fragility behavior in Ge-V-VI (V=As, or Sb, and VI=S, Se, Te) alloys prove to be very different and the differences can be rationalized in terms of the known electronic conductivities and SC-M transitions, thereby connecting them to the differences in metallicity of the individual alloy components. Finally, we will consider how liquid fragility transitions associated with the SC-M transitions might influence the crystallization kinetics in PCM materials and thereby account for the unique combination of ultrafast phase switching with large changes of electrical conductivity that characterize PCMs.

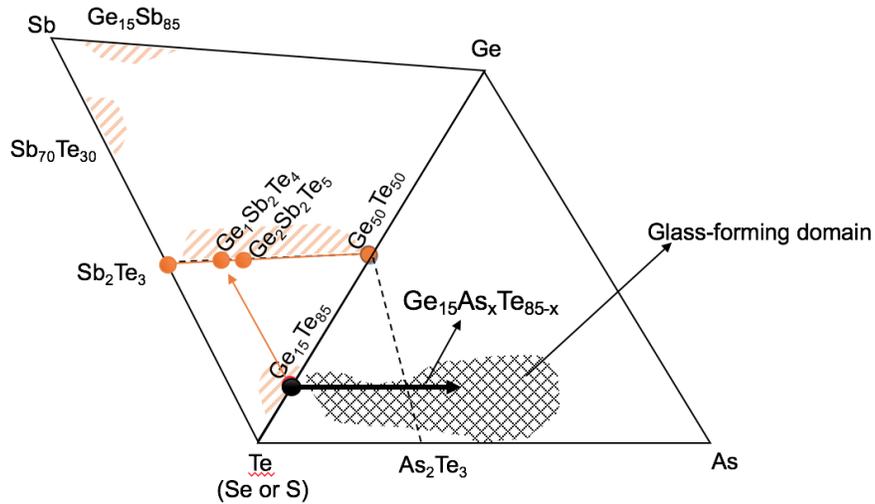

**Figure 1.** Ternary phase diagrams of Ge-As-Te and Ge-Sb-Te with common Ge-Te boundary. In the Ge-As-Te ternary diagram (right triangle), the arrows indicate the composition line of $Ge_{15}As_xTe_{85-x}$ alloys studied in this work. The black shaded area is the glass-forming domain determined by the water quenching method according to ref.[27]. In the Ge-Sb-Te ternary diagram (left triangle), the orange dots and lines mark typical compositions of phase-change materials

according to ref. [2]. The thin orange arrow indicates the corresponding compositions in the Ge-Sb-Te system that we sought to explore, with only limited success due to fast crystallization.

## II. MATERIALS AND METHODS

Amorphous $Ge_{15}As_xTe_{85-x}$ and $Ge_{15}Sb_2Te_{83}$ samples were prepared using the Ge, As (or Sb) and Te elements with purities ranging from 99.999 to 99.9999 at. %, sealed under vacuum ($10^{-6}$ mbar), and synthesized in a rocking furnace for homogenization at 900 °C for 15 hrs. The melt was subsequently quenched sidewise into ice-water to achieve a high cooling rate, resulting in a ~1 mm amorphous layer. For compositions $Ge_{15}Sb_xTe_{85-x}$ where x>2, fully amorphous samples could not be obtained using ice-water quenching or a technique with a higher quenching rate using ice + ethanol bath, due to the poor glass-forming ability caused by increased Sb-concentration.

Differential Scanning Calorimetry (DSC) studies were carried out using a TA Ins. Q1000 MDSC (TA DSC) for fragility determinations, which were carried out near $T_g$ and a Perkin Elmer Diamond DSC (PE DSC) is used to extend measurements to higher temperatures near and above 500 °C. Amorphous samples with mass about 5-15 mg were sealed in aluminum T-Zero pans, and an empty aluminum pan was used as a reference. TA DSC cell calibration was carried out with standard sapphire. Both temperature and enthalpy were calibrated for each heating rate using standard indium (and zinc) prior to measurements for TA DSC (and PE DSC). The error of measured temperature is within 0.5 K.

Fragilities for a series of compositions along the thick black arrow of Figure 1 were determined using the cooling rate dependence of the fictive temperature method, and also instrumentation (TA DSC), described in our previous work[21,22] on the fragility of $Ge_{15}As_xSe_{85-x}$, $Ge_{15}As_xS_{85-x}$, and also as one of the methods employed in the DSC study of $Ge_{15}Te_{85}$[23]. The fictive temperature is defined, for a given cooling rate, $q_c$, as the temperature of onset of the heat capacity jump during upscan at the same rate i.e. $q_h = q_c$, since the identity $T_f \approx T_g^{onset}$ was shown in earlier work[28] to hold under this protocol. More specifically, DSC scans were carried out from 25 °C through the glass transition until well above $T_g$ to ensure the system reached the metastable equilibrium of the supercooled liquid state before cooling at the next $q_c$ for the next run (for cases where crystallization at T>$T_g$ was unlikely). $T_g^{onset}$ is determined by the usual tangent construction illustrated in the lowest curve of Figure 2.

Figure 2 shows a series of scans for $T_f$ determinations at different q values ranging from 3 to 30 K min$^{-1}$ for a single composition (Ge$_{15}$Sb$_2$Te$_{83}$). For the poorer glassformers, (e.g. Figure 2), each scan was obtained with a fresh sample as a precaution against any crystallization that might occur during the short exposure to T > $T_g$. A series of scans, like that of Figure 2, yields the value of fragility for a single composition, by plots described in the Results section.

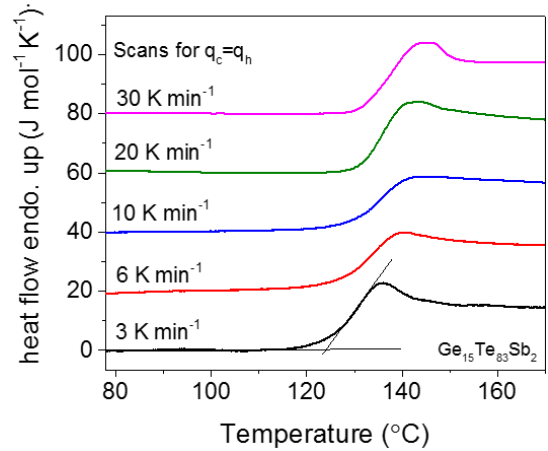

**Figure 2.** DSC heat flow of Ge$_{15}$Sb$_2$Te$_{83}$ measured upon heating at various rates $q_h$ after previous cooling ($q_c$) from supercooled liquid at the same rate $q_c = q_h$. The onset temperatures of the glass transitions are determined by tangent line constructions as illustrated for $q_h$ = 3 K min$^{-1}$. The curves are vertically shifted for clarity.

## III. RESULTS

Figure 3 shows the DSC heat flow during heating at 20 K min$^{-1}$ of as-quenched amorphous samples for compositions Ge$_{15}$As$_x$Te$_{85-x}$ along the line illustrated in Fig. 1 and Ge$_{15}$Sb$_2$Te$_{83}$. Glass transitions $T_g$ are observed as endothermic events on the DSC scans (marked by solid triangle). With increasing temperature above $T_g$, the supercooled liquids crystallized (indicated by exothermic peaks - $T_x$ marks the onset of crystallization), followed by endothermic peaks of melting that occur over a range of temperature (arrows point to the onset $T_e$ and liquidus ($T_L$) temperatures). TABLE I lists the values of these transition temperatures, enthalpies of fusion Δ$H_m$ and of crystallization Δ$H_x$ derived from integrations of heat flow peaks. Figure 4 shows that $T_g$ of $Ge_{15}As_xTe_{85-x}$ increases with increasing As concentration; whereas $T_e$ displays weak changes as the melting processes may involve the mixtures of different metastable phases formed during crystallization upon heating. $T_L$ becomes higher as As-atoms replace Te-atoms in the alloys. Figure 4(b) shows the Turnbull parameter[29] $t_{rg}=T_g/T_L$, which is commonly used as an indicator of thermodynamic driving force for crystallization, and the parameter $\Delta T_{x-g}=T_x-T_g$. The latter is a measure of the width of the supercooled region that is accessible to experiment.

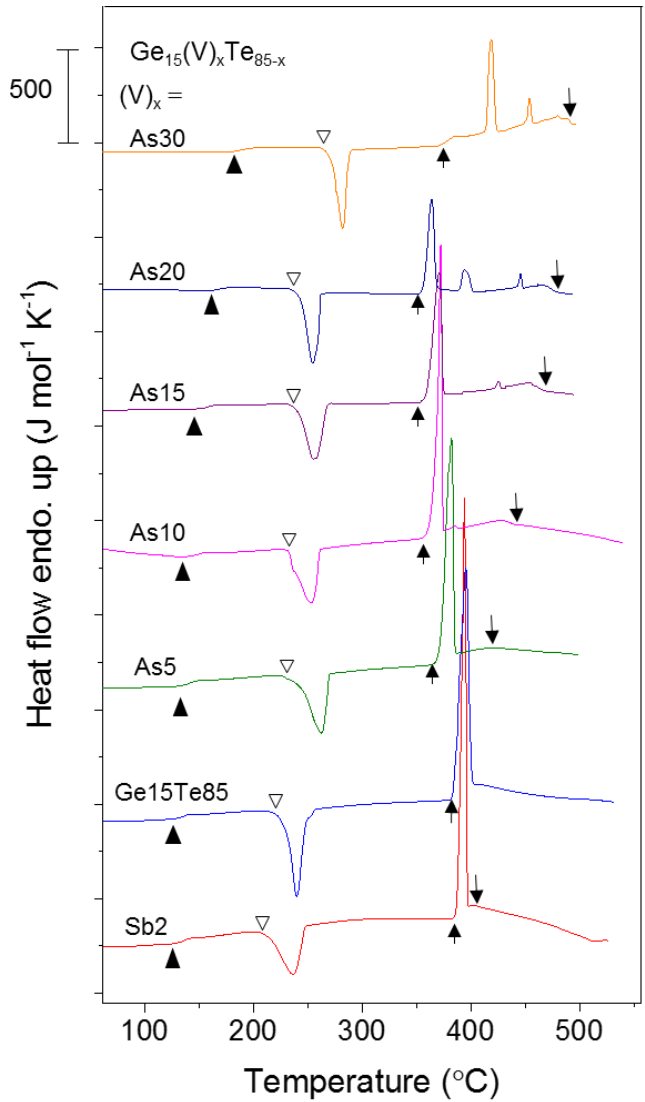

**Figure 3.** DSC heat flow upon heating at 20 K min$^{-1}$ for as-quenched samples of $Ge_{15}As_xTe_{85-x}$ and $Ge_{15}Sb_2Te_{83}$. Solid and open triangles indicate the glass transition temperatures $T_g$ and crystallization onset temperature $T_x$, respectively. The arrows point to the onset ($T_e$) and the end ($T_L$) of melting. The composition As30 apparently belongs to a different phase field. The excess heat flow for the binary (starting) composition $Ge_{15}Te_{85}$, and $Sb_2$ seen on the high temperature side of the eutectic temperatures is due to the residuum of the liquid state $C_p$ anomaly which was the focus of our previous article[23]. The composition line under study would appear to be closely following a eutectic valley from $Ge_{15}Te_{85}$ down to a ternary eutectic near 20% As (in Fig. 1)

**TABLE I.** Summary of $T_g$, $T_x$, and $T_L$ of as-quenched samples measured at 20 K min$^{-1}$. The activation energy $E_a$ at $T_g$ and the fragility $m$ are determined using DSC methods. $\Delta T_{x-g}$ are average values from three measurements for as-quenched samples. The enthalpy of crystallization $\Delta H_x$ and enthalpy of fusion $\Delta H_m$ are determined by integrating exothermic crystallization and endothermic melting peak areas, respectively. Estimated uncertainties are included or implied by significant figures.

|  | $T_g$ (K) | $T_x$ (K) | $T_L$ (K) | $E_a$ (kJ mol$^{-1}$) | Fragility $m$ | $\Delta T_{x-g}$ (K) | $t_{rg}$ | $\Delta H_x$ (kJ mol$^{-1}$) | $\Delta H_m$ (kJ mol$^{-1}$) |
|---|---|---|---|---|---|---|---|---|---|
| Ge$_{15}$As$_{30}$Te$_{55}$ | 454.0 | 538 | 766 | 261+12 | 30±3 | 84 | 0.593 | 4.3 | 7.0 |
| Ge$_{15}$As$_{20}$Te$_{65}$ | 433.9 | 510 | 756 | 227±8 | 27±1 | 76 | 0.574 | 4.4 | 7.2 |
| Ge$_{15}$As$_{15}$Te$_{70}$ | 424.5 | 505 | 743 | 229±15 | 28±2 | 81 | 0.571 | 4.9 | 9.7 |
| Ge$_{15}$As$_{10}$Te$_{75}$ | 411.0 | 505 | 714 | 277±14 | 35±2 | 94 | 0.574 | 5.2 | 9.5 |
| Ge$_{15}$As$_5$Te$_{80}$ | 406.3 | 498 | 723 | 324±26 | 41.5±3 | 93 | 0.562 | 5.6 | 11.4 |
| Ge$_{15}$Te$_{85}$ | 403.1 | 475 | 658[a,b] | 380+28 | 49[b]±3 | 89 | 0.613 | 5.9 | 10.6 |
| Ge$_{15}$Sb$_2$Te$_{83}$ | 403.7 | 482 | 679 | 394±21 | 51±3 | 84 | 0.595 | 5.1 | 10.2 |

[a] eutectic temperature

[b] ref. 23

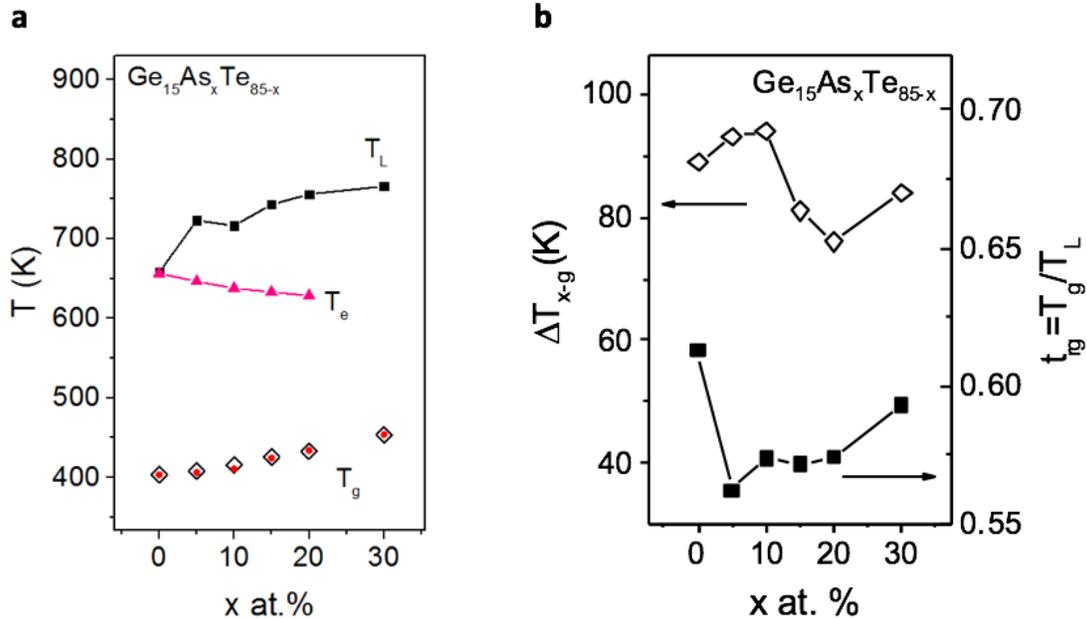

**Figure 4.** (a) Glass transition $T_g$, melting onset (or eutectic) $T_e$ and liquidus $T_L$ temperatures of Ge$_{15}$As$_x$Te$_{85-x}$ as a function of $x$ values. The open diamonds represent $T_g$ for as-quenched glasses measured at 20 K min$^{-1}$ while dots (inside diamonds) are standard $T_g$ values under the constraint

$q_h=q_c$=20 K min$^{-1}$. **(b)** The Turnbull parameter $t_{rg}$ (solid squares) and the width of the supercooled region $\Delta T_{x-g}$ (open diamonds). Uncertainties on the individual points are comparable with the symbol sizes.

=====================

Next we characterize the temperature dependence of structural relaxation times (i.e. fragility) near the glass transition. The DSC scans are carried out at a series of cooling/heating rates $q \equiv q_h = q_c$ throughout the glass transition from room temperature to well above $T_g$ for each composition using the procedures described in the Experimental Section. The obtained fictive temperatures $T_f$ and corresponding heating (=cooling) rates $q$ for Ge$_{15}$As$_x$Te$_{85-x}$ and Ge$_{15}$Sb$_2$Te$_{83}$ are plotted in Arrhenius form in Fig. 5. From the commonly used plot of [$ln(q)$ vs. $1/T_f$], the activation energy $E_a$ near $T_g$ can be determined by calculating the slope of a linear fit, since the slope is equal to -$E_a/R$. The $E_a$ is related to fragility by $m=E_a/(ln10 \cdot T_g \cdot R)$, (ref.[8]), where $T_g$ is the glass transition temperature measured using a "standard DSC scan" (i.e. upscan at 20 K min$^{-1}$ after cooling at the same rate, $q_h=q_c$=20 K min$^{-1}$). $T_g$ defined this way corresponds to the temperature at which the structural relaxation time $\tau$~100 s [30]. The results of $E_a$ and the fragility $m$ are summarized in TABLE I. As shown in Fig. 6, the fragility of Ge$_{15}$As$_x$Te$_{85-x}$ is initially lowered by a small amount of As, and reaches the lowest $m$ value, 27, at around x=20, i.e. the strongest liquid behavior is observed at Ge$_{15}$As$_{20}$Te$_{65}$. With further increase of As concentration, the system tends back to higher fragilities. Such a fragility behavior resembles that of Ge$_{15}$As$_x$Se$_{85-x}$ and Ge$_{15}$As$_x$S$_{85-x}$ (ref.[21,22]) and exhibits almost the same <r>- and As at. %-dependence**.**

By striking contrast, replacement of As by Sb as the third component at the same constant Ge causes $m$ to change in the opposite direction. It is unfortunate that fast crystallization excluded the exploration of this effect beyond the composition Ge$_{15}$Sb$_2$Te$_{83}$. More Sb-rich compositions

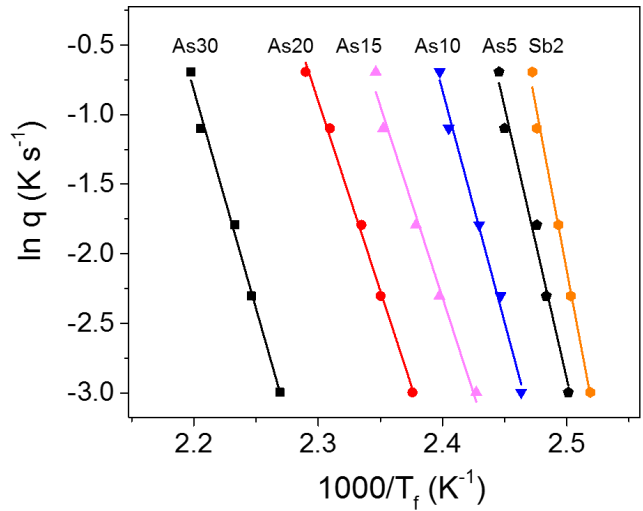

**Figure 5.** The natural logarithm of scanning rate, *ln q*, where $q=q_h=q_c$ holds, is plotted against $1/T_f$ for $Ge_{15}As_xTe_{85-x}$ (x=30, 20, 15, 10, 5) and $Ge_{15}Sb_2Te_{83}$. The slope of the fitted line is equal to -$E_a/R$, which is related to fragility by $m=E_a/(ln10 \cdot T_g \cdot R)$, where $T_g$ is the glass transition temperature measured under a standard DSC scan, $q_h=q_c=20$ K min$^{-1}$. Note that the same *m* values can be also obtained by constructing a [*log q/q_s* vs $T_f^s/T_f$] plot (see details in ref.[23,30]).

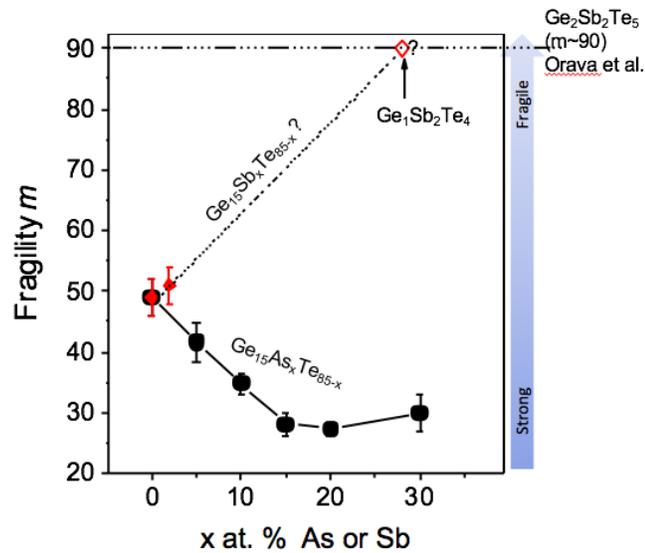

**Figure 6.** The fragility *m* of $Ge_{15}As_xTe_{85-x}$ (solid circles) liquids as function of at % As. *m* ranges from intermediate to strong with increasing As content. For $Ge_{15}Sb_xTe_{85-x}$, fragility data are only available for x=2 (solid diamonds). The open diamond represents the fragility m~90 for x=28 (~ GST124, see text for details) by assuming that GST124 has a similar fragility to that of GST225 (m~90) estimated by Orava et al.[4].

================================================

were found always to be contaminated by crystalline material. While this is certainly consistent with the proposed high fragilities in the Sb-containing ternaries, our composition range is too small to serve as a confirmation. The dotted line extrapolating our finding to the domain of the PCM GST124 is barely scientific. However if we assume that GST124 has a fragility similar to the value $m \approx 90$ estimated for the neighboring GST225 from crystallization kinetics[4], then GST124 (which is $Ge_{15}Sb_xTe_{85-x}$ with x=28) can be represented by the open diamond in Fig. 6, where it supports the dramatic difference between the As-based and Sb-based systems with respect to fragility behavior. Such a contrast in the effect of the group V element on the physical behavior is certainly eye-catching and its physical origin demands an explanation. This is taken up in the following Discussion section after some consideration of the glass transition temperatures themselves, in relation to chalcogenide component.

## IV. DISCUSSION

The increase of $T_g$, as crosslinking elements like Ge and As are added to the chalcogen selenium, is a familiar phenomenon[20,31,32]. Not so well recognized is the more rapid increase in the case of the chalcogen sulfur[22] and, of special importance to us, the much *less* rapid increase when the chalcogen is Te. The relations are summarized in Figure 7. It has been common, following Phillips[33,34] and Thorpe[35] to use the bond coordination number, or bond density, $<r>$, as the correlating composition variable, although more recent studies[21] suggest it is an oversimplification. $<r>$ is obtained by summing the products of bond number (from the 8-N rule) and atom fraction in the sample. Thus for a Ge-As-Te glass with 15% Ge, it is given by $<r> = [4 \times 15 + 3 \cdot x + 2 \times (85-x)]/100$, where $x$ is the atom percentage of As in the sample. We notice in Figure 7 that, while the increase in $T_g$ is rapid when the chalcogen that is being crosslinked by the As or Ge is S or Se, when it comes to Te, the increase in $T_g$ is small. Finally, when As is replaced by Sb, as in the PCMs, the $T_g$ increase, to the extent that it can be measured or estimated, virtually disappears for $<r>$ greater than 2.3, which is the crossover value, where all systems have $T_g$ of 400 K. For instance, Kalb et al. studying sputtered compositions that are annealed before scanning to remove quenched-in high temperature structure, find $T_g = 411$ K for GST124 (with $<r> = 2.57$) and 430 K for GST225 ($<r>$=2.67). On the other hand, Orava et al.[4] found reasons for using the much lower value of $T_g$ =383 K for GST225. The confused state of debate is reminiscent of the notorious case of vitreous water, where hyperquenched glass studies also indicate high $T_g$ values[36–38], while work on glasses formed by other routes[39,40], or extrapolated from binary solutions[41,42], yield much lower values. In any case it appears that covalency of interaction, and the related average bond density, can no longer be the dominant consideration for GST glasses.

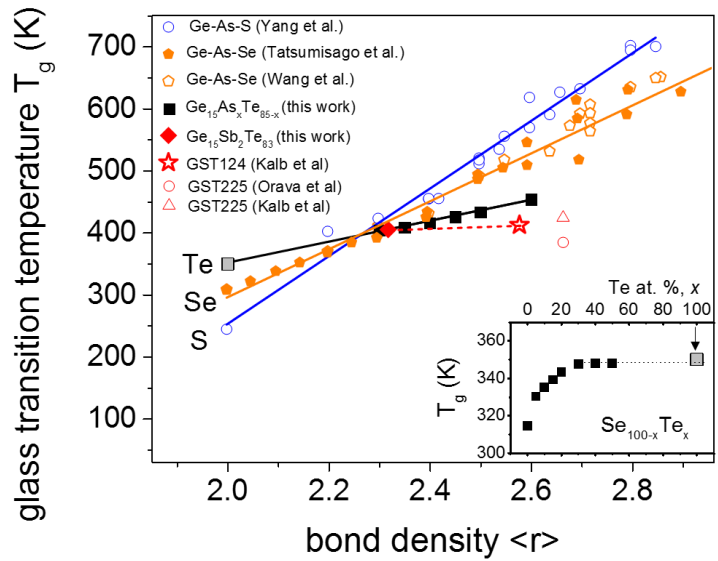

**Figure 7**. Glass transition temperature $T_g$ as a function of bond density $<r>$ for $Ge_{15}As_xTe_{85-x}$ (this work), Ge-As-Se, (Tatsumisago et al.[20] and Wang et al.[21]), and Ge-As-S (Yang et al.). The linear variation is a good approximation up to $<r>$ values of 2.4. Beyond $<r> = 2.4$, composition-specific effects are found[20,22], causing the spread of data points. More scattered data for Ge-As-Te glasses from early work by Savage[43] and additional data from Lucas et al.[44] (not shown) support the present findings at slightly lower $T_g$ values. For Te- based glasses with $<r>$ increased by Sb instead of As (thus entering the fast-crystallizing PCM domain) the behavior becomes uncertain and confusing. Some estimates imply weak positive dependence of $T_g$ on $<r>$ while others require decreases, as discussed in text. **Inset:** the glass transition temperature of $Se_{100-x}Te_x$ alloys[45]. The grey square is the extrapolated $T_g=350$ K of pure Te and is consistent with the value extrapolated from the data in the main panel.

======================================================================

Indeed, when the electronic conductivity of these materials is examined it becomes clear that the concept of electron localization in a covalent bond is becoming very fuzzy. Figure 8, adapted from the works of Alekseev et al.[46] and Nagels et al.[47] presents electronic conductivity data in Arrhenius form for a variety of chalcogenides as they pass from low temperature semiconducting states to high temperature "weak metal" states (at the Mott minimum metallic conductivity, ~ $10^3$ $\Omega^{-1}cm^{-1}$, through a maximum in the apparent activation energy (Figure 8a)).  Figure 8b shows how this semiconductor-to-metal (SC-M) transition, at 1280 K for the case of $As_2Se_3$, occurs at the same temperature as the closing of the optical band gap (1250 K), which is also the temperature of a thermodynamic anomaly in the density (density minimum and maximum with thermal expansion coefficient extremum, $\alpha$(min), at ~1280 K) ($\alpha=1/V(\partial V/\partial T)_p$). The transition occurs far above the melting point of 641 K[48] for $As_2Se_3$, but when Se is replaced by Te the density anomaly, with extremum $\alpha$(min), occurs at 780 K, much closer to $T_m$ (640K)[49]. Finally, when the more metallic bismuth replaces antimony, the liquid $Bi_2Te_3$ is already metallic at its 857 K[50] melting point (see Fig. 8c), meaning the SC-M transition has been pushed below the melting point and can no longer be observed due to fast crystallization. If the SC-M transition is sharp enough, then the accompanying heat capacity change will be sharp, and according to Adam-Gibbs theory, a fragile-to-strong transition (in which the viscosity changes more or less suddenly by some orders of magnitude) will occur during cooling[23] . The importance of this is discussed further below.

Considering lower temperatures in the glassy state, one can see how dramatically the glassy conductivity responds to the change in metallicity of the components[47]. Figure 8c shows the effect of doping $GeSe_{3.5}$ glasses alternatively with just 12 at % Sb vs. 12 % Bi in the system $(GeSe_{3.5})_{88}Sb_{12}$, can cause a change of some three orders of magnitude.

As the band gap is closing, an increasing fraction of the valence electrons involved in the covalent bonds are promoted into the conduction band, or into the more diffuse tail states of Mott Anderson theory[51], where they are delocalized, and the constraint counting involved in the assessment of average bond density $<r>$ becomes increasingly irrelevant to the physics of the system. For this reason, it is expected that the dependence of $T_g$ on $<r>$ should become weaker to vanishing. The latter is the case of Ge-Sb-Te phase-change alloys.

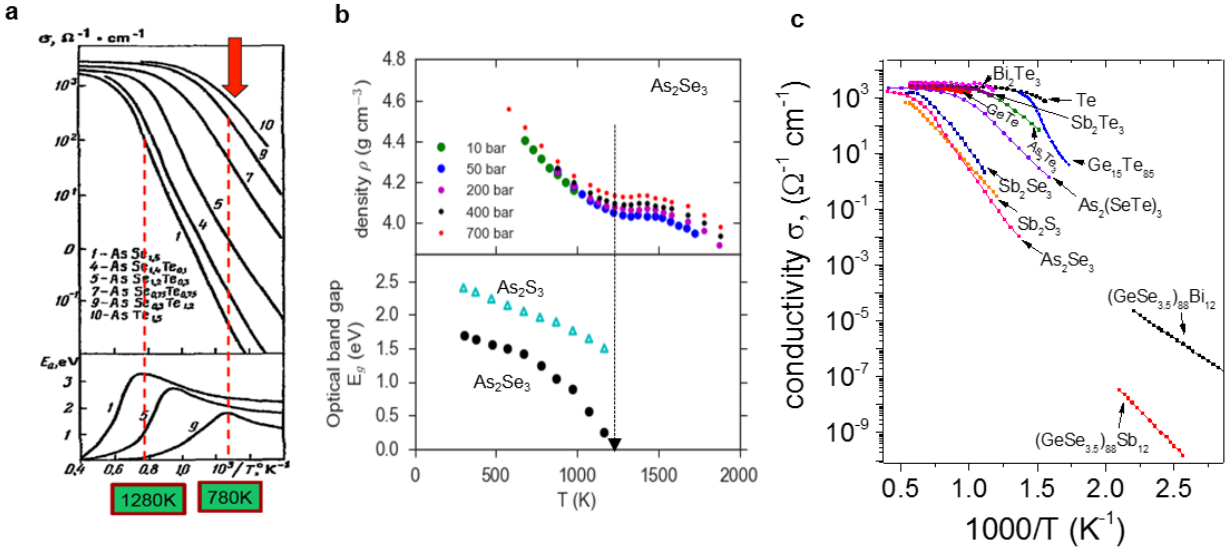

**Figure 8**. (**a**) Electronic conductivity of liquid $As_2Se_3$ – $As_2Te_3$ alloys (upper panel) and the apparent activation energy $E_a$ for conductivity (lower panel) (adapted from ref. [46]). The dashed lines mark the temperatures of the maximum $E_a$ i.e. the SC-M transition temperature $T_{SC-M}$. The arrow points to the temperature $T(C_p^{max})$ of the heat capacity anomaly of $As_2Te_3$ (780K) (ref.[49]). (**b**) Upper panel: Densities of liquid $As_2Se_3$ at various pressures showing density anomalies. The phase transition temperature is where the minimum in $\alpha(T)$ occurs. Lower panel: Optical band gap of liquid $As_2Se_3$ closes at ~1250 K, and that of $As_2S_3$ closes at a higher temperature. Data taken from ref.[52,53]. (**c**) Electronic conductivities for a variety of liquid chalcogenides, and two glasses (data reproduced from ref.[46,47]). Note that the conductivities approach a plateau of the order of magnitude ~$10^3$ $\Omega^{-1}$ $cm^{-1}$ (the Mott minimum metallic conductivity) by a SC-M transition.

==========================================================

The decreasing relevance of the bond density is also reflected in the loss of glass-forming ability. For instance, in the case of the composition $As_2Se_3$ for which $<r>$ = 2.4 (the "magic" bond density), the "good glassformer" characteristics were used as support for the constraint-counting theory[33]. But when there is total replacement of Se by Te to give the more metallic $As_2Te_3$, the $<r>$ value remains the same but the glassforming ability is lost. In that case the SC-M transition, at 780 K, has closely approached the melting point of 640 K, and it is clear that if the As is replaced by the next lower in Group V, Sb, the temperature of the SC-M transition will fall below the melting point. Indeed, as seen in Figure 8c, $Sb_2Te_3$ is metallic at its melting point. The problem is then to decide how far below the melting point the transition actually would occur, what could be the physical consequences, and finally to what extent a substantial Ge component in GST PCMs would modify the effect of Sb replacement of As in the series just discussed. A study of the liquid-liquid (LL) transition with increasing Sb in the binary system $As_2Te_3$-$Sb_2Te_3$ would be helpful, but does not appear to have been made. We would predict that systematic changes in the temperature of the SC-M transition would be observed as Sb content increases, and that an extrapolation could be made to locate the sub-$T_m$ value for $Sb_2Te_3$.

At this point it is profitable to consider the phenomenologically related case of the "most anomalous" liquid, viz. water and particularly supercooled water, for which a strict limit on supercooling is found at a temperature that is some 15% below its melting point. This is apparently a direct consequence of the structural fluctuations associated with an impending LL phase transition. The phenomenology of "hidden" LL phase transitions, and their relation to fast crystallization, in supercooled water has been the subject of exhaustive studies both by experiment and computer simulation[54], to which brief reference was made in our preceding paper[23]. In the water case, the response functions heat capacity, compressibility, and expansivity of water all show highly anomalous behavior with an apparent common divergence temperature just below the sharply defined homogeneous nucleation temperature (which can be measured for aqueous systems using small sample techniques). Furthermore, the liquid immediately above the supercooling limit is extremely fragile in character. To date it has still not been possible to reach clear conclusions on whether the transition in the laboratory substance $H_2O$ at ambient pressure is first order as in liquid Si[55], or higher order, as in $As_2Te_3$. To add plausibility to this parallel, we note the recent structural studies on GST124 by Clark and coauthors[56], that not only lead to observation of high pressure amorphization of the crystalline form, (analogous to the pressure induced amorphization of ice Ih by Mishima[57], but give evidence for a reversible polyamorphic phase transition when starting with the sputtered glassy form, comparable to the famous studies of Mishima for glassy water[58].

If we translate these observations to the liquid GST case (specifically GST225), we can begin to understand the reason that crystallization studies have yielded the conclusion that the crystallizing liquid is very fragile ($m \approx 90$ ref.[4]). It should be because they are being studied on the high temperature side of the LL (now also SC-M) transition. To gain some additional support for this notion, we can plot the temperatures of the heat capacity maxima (i.e. the SC-M transition temperatures) on the phase diagram for Ge-Te and extrapolate to the PCM composition GeTe. This is shown in Figure 9, where it is seen crudely to locate the SC-M transition at ~ 800 K, some 20% below the melting point (cf. the 15% for water, noted above).

Finally, to support this thesis, we would note the work of Zalden et al on the $Ge_{15}Sb_{85}$ alloy, another PCM that is used in phase-change memory devices. Using short laser pulses, Zalden et al.[59] induced fast melt-quench cycles in $Ge_{15}Sb_{85}$ and employed ultrafast X-ray scattering to probe the structural changes of the undercooled state. A structural transition was observed just before crystallization sets in, although it is difficult to determine the exact temperature of the transition. Supporting this observation, and our arguments, is also a very recent ab initio calculation on GST225 that reported a pseudo-gap at the Fermi energy which is opening with decreasing temperature, suggesting a SC-M transition would occur at a somewhat lower temperature, below 852 K[60].

Thus the connection between submerged LL phase transitions, fragile liquid behavior and fast crystallization kinetics, gains some credibility, without being established. From water phenomenology, the LL transition should always occur below the temperature of the density maximum. A more detailed investigation of this phenomenology, and indirect means of assessing the relation between SC-M and melting transition temperatures, will be the subject of a broad overview paper to be published separately.

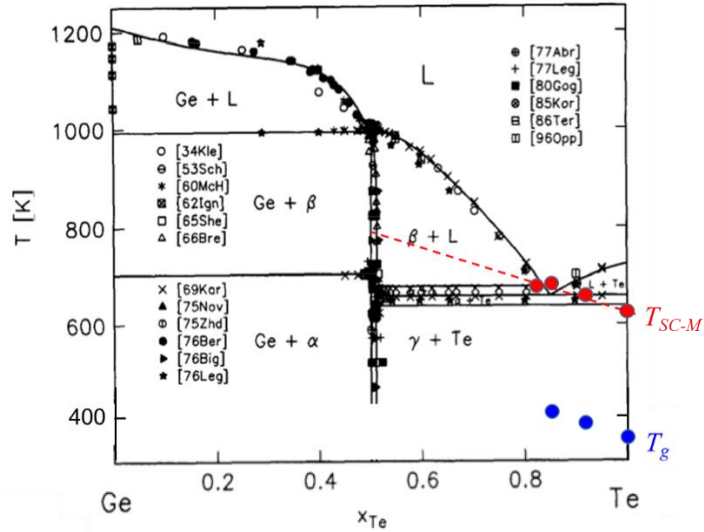

Figure 9. Preliminary estimate of the SC-M transition temperature, $T_{SC-M}$ in a phase-change material GeTe on the binary Ge-Te phase diagram[61]. An extrapolation to ~800 K for GeTe is based on the temperatures of liquid $C_p$(max) (red solid circles) (ref.[62,63]). $T_g$ are represented by solid blue circles ($Ge_{15}Te_{85}$: ref.[23], $Ge_{10}Te_{90}$: ref.[64], and Te: 350 K from Fig. 7).

==============================================

**Synopsis and technological relevance**

For phase-change memory applications that greatly improve on the original and most common "flash" memory devices, PCMs must possess a unique combination of properties, namely, extremely fast phase switching (timescale of nanoseconds) at high temperature, and a high stability of amorphous and crystalline phases at room temperature, as well as a strong optical/electrical contrast between the two phases. The optical/electrical contrast owes its existence to the metal-semiconductor transition, which we have particularly noted is closely linked to structural fluctuations that are jointly the source of thermodynamic anomalies and enhanced nucleation rates (as has been so well studied in supercooled water). If these can be induced to occur at an optimum supercooling, (i.e. optimum $T_{SC-M}/T_m < 1$), they should facilitate the crystallization process and greatly speed up the phase switching, thus explaining how nanosecond switching behavior is possible in PCMs. But when cooling rate is increased sufficiently to avoid crystallization the fragile-to-strong liquid transition that is companion to the M-SC transition will quickly lock in the amorphous configuration in a kinetically stable vitreous state until a new heat pulse raises it back to the fast crystallization domain.

The applied technology problem, then, is to ensure that, on the one hand $T_{SC-M}/T_m$ is low enough for the thermodynamic crystallization driving force $G_{liq}-G_{cryst}$ to be sufficient to drive the process, while on the other hand, that $T_{SC-M}/T_m$ is not so low that the system becomes a viscous glassformer that retains the metallic state and so shows little electrical contrast even if crystallization should occur. Accordingly, a key to the favored properties of PCMs is the "tuning" of $T_{SC-M}/T_m$ by, for instance, doping in of elements of the appropriate metallicities whose role in moving the SC-M transition has been discussed above.

## V. Concluding Remarks.

It is intriguing to find such high levels of phenomenological similarity between such seemingly different condensed phases of matter as we have suggested in the foregoing discussion. However, this is not a new observation, except for the connection to PCM phenomenology. The extraordinary parallel in physical behavior between water and the element tellurium in their supercooled liquid states was detailed in the $T_m$-scaled plots of volume, heat capacity, and isothermal compressibility by Kanno et al.[65] in 2001, and interpreted by one of us[66] as a consequence of each liquid being characterized by two different length scales - as in the Jagla model of anomalous liquids[67,68]. There are frequent references in the PCM literature to the two different Ge coordination states that might play an equivalent role.

While there are controversies associated with the fact that water and Te are most anomalous in their supercooled states[69], the LL phase transition in the starting composition for our study, $Ge_{15}Te_{85}$, has its heat capacity maximum in the thermodynamically stable domain just above the melting point (eutectic temperature) ($T_{SC-M}/T_m \approx 1.01$) and thus cannot be regarded as a transient Ostwald stage on the route to crystallization, as has been proposed for water[69]. This transition and this composition can then be seen to lie at the crossover between the two sorts of behavior, i.e. at the point, at which covalency loses its control of structure in favor of metallicity. And then the fluctuations in structure, which accompany the transition, enhance the probability of crystal nucleation in proportion to the degree of supercooling at which they reach their maximum value. The possibility that a nearby SC-M transition may facilitate crystallization and speed up the phase switching implies, as already stated above, that a key parameter in the assessment of PCM phenomenology should be the temperature relative to the melting point, $T_{SC\text{-}M}/T_m$, at which the SC-M transition occurs. However, this reduced temperature $T_{SC\text{-}M}/T_m$ must be very difficult to observe in PCMs by direct measurement. As in the case of water, it will need to be assessed by measurements made in or near the stable liquid domain, (preferably in microscopic samples to reduce nucleation probabilities), followed by fitting to appropriate theoretical functions. It will likely remain unknown whether or not the SC-M transition is first order in some cases or continuous as it is in the examples of Figure 8. Ab initio computer simulations will have an important role to play in clarifying the outstanding issues. The possibility of a liquid-liquid critical point playing a role was conjectured long ago[70].

## ACKNOWLEDGEMENTS

The authors acknowledge NSF collaborative research Grant Nos. CHE 12-13265 and ECCS-1201865. S.W. acknowledges the financial support from an Alexander von Humboldt-Foundation Feodor Lynen Postdoctoral Research Fellowship.